\documentclass[aps,prl,showpacs,twocolumn,superscriptaddress]{revtex4}

\usepackage[latin1]{inputenc}
\usepackage[T1]{fontenc}
\usepackage{times}
\usepackage{tabularx}
\usepackage{amsmath,graphicx}

\newcommand{\ket}[1]{| #1\rangle} 

\newcommand{\ic}{\mathrm{i}}

\begin{document}

\title{Two scenarios for quantum multifractality breakdown}

\author{R. Dubertrand}
\affiliation{%
Universit\'e de Toulouse; UPS; Laboratoire de
 Physique Th\'eorique (IRSAMC); F-31062 Toulouse, France
}
\affiliation{CNRS; LPT (IRSAMC); F-31062 Toulouse, France}
\author{I. Garc\'{\i}a-Mata}
\affiliation{Instituto de Investigaciones F\'isicas de Mar del Plata
(IFIMAR), CONICET--UNMdP,
Funes 3350, B7602AYL
Mar del Plata, Argentina.}
\affiliation{Consejo Nacional de Investigaciones Cient\'ificas y
Tecnol\'ogicas (CONICET), Argentina}
\author{B. Georgeot} 
\affiliation{%
Universit\'e de Toulouse; UPS; Laboratoire de
 Physique Th\'eorique (IRSAMC); F-31062 Toulouse, France
}
\affiliation{CNRS; LPT (IRSAMC); F-31062 Toulouse, France}
\author{O. Giraud}
\affiliation{LPTMS, CNRS and Universit\'e Paris-Sud, UMR 8626, B\^at. 100,
91405 Orsay, France}
\author{G. Lemari\'e}
\affiliation{%
Universit\'e de Toulouse; UPS; Laboratoire de
 Physique Th\'eorique (IRSAMC); F-31062 Toulouse, France
}
\affiliation{CNRS; LPT (IRSAMC); F-31062 Toulouse, France}
\author{J. Martin}
\affiliation{Institut de Physique Nucl\'eaire, Atomique et de
Spectroscopie, Universit\'e de Li\`ege, B\^at.\ B15, B - 4000
Li\`ege, Belgium}
\date{February 7, 2014}

\begin{abstract}
We expose two scenarios for the breakdown of quantum multifractality under the effect of perturbations. In the first scenario, multifractality survives below a certain scale of the quantum fluctuations.  In the other one, the fluctuations of the wave functions are changed at every scale and each multifractal dimension smoothly goes to the ergodic value. We use as generic examples a one-dimensional dynamical system and the three-dimensional Anderson model at the metal-insulator transition.  Based on our results, we conjecture that the sensitivity of quantum multifractality to perturbation is universal in the sense that it follows one of these two scenarios depending on the perturbation. We also discuss the experimental implications.
\end{abstract}
\maketitle

The notion of multifractality is associated to scale-invariant fluctuations which cannot be described by a single fractal dimension but instead by an infinite number of dimensions. This property characterizes several important systems in classical physics, e.g. turbulence \cite{turbulence}, the stock market \cite{stock}
or cloud images \cite{cloud}. It is only recently that multifractality has been recognized in quantum mechanics or other wave systems.   Examples include
electrons at the Anderson
metal-insulator transition \cite{kohmoto,mirlin2000,mirlinRMP08,romer}, 
quantum Hall transitions \cite{huckenstein}, Random Matrix models \cite{PRBM,ossipov} and others \cite{indians1,indians,garciagarcia}.  These properties are also visible in the wave functions in certain types of dynamical systems (so-called pseudointegrable systems) \cite{interm,wiersig,giraud,bogomolny,MGG,map_pseudoint,MGGG,BogGir,MGGG12}. 

Although many theoretical studies have been devoted to quantum multifractality \cite{kohmoto,mirlin2000,mirlinRMP08,romer,huckenstein,PRBM,ossipov,indians1,indians,garciagarcia,interm,wiersig,giraud,bogomolny,MGG,map_pseudoint,MGGG,BogGir},  in particular to its dynamical consequences which seem accessible to experiments \cite{guarneri,geisel,mantica,MGGG12}, it has been difficult to observe it in a real setting. Hints of such properties were seen 
in disordered conductors \cite{richard} and cold atoms \cite{cold,cold2}. An interesting experiment enabled to observe multifractal distributions at the Anderson transition with acoustic waves \cite{billes}.  
However, experimental characterization of multifractality in a quantum context has remained elusive.  There are technical questions related to the high resolution needed to explore different scales in the wave function, but fundamentally it is of critical importance to assess to what extent multifractality survives in a real experimental setting. 

In this Letter, we study the effects of different imperfections and perturbations on the properties of two paradigmatic models with quantum multifractality, a one-dimensional (1D) dynamical system and the three-dimensional (3D) Anderson model at the metal-insulator transition.
We find that a sufficiently large perturbation always destroys multifractality, but in two different ways. In the first scenario, the perturbation defines a new scale of the quantum fluctuations below which multifractality survives.  In the second scenario, the fluctuations of the wave functions are changed at every scale and each multifractal dimension smoothly goes to the ergodic value.  Our results show that both scenarios are found in the two models, depending on the type of perturbation.

Multifractality of quantum wave functions $ | \psi \rangle$ can be characterized by the box-counting method (see \cite{romer,MGGG} for comparison with other methods). A system of linear size $L$ is divided into $L/\ell$ boxes of size $\ell$, and a measure for each box $k$ is $\mu_k=\sum_i |\psi_{i}|^2$ where the indices run over the sites inside box $k$.  The moments are defined by $P_{q}=\sum_{k} \mu_k^q$.
Multifractality is characterized by a power-law behavior of the moments $P_{q} \sim (\ell/L)^{D_q (q-1)}$, 
in the limit of small $\ell/L$. In the ergodic limit all $D_q$ equal the dimensionality of the system, whereas for a localized system $D_q=0$ for $q>0$. In
systems where an average is made over several wave
functions and different disorder realizations, two sets of multifractal dimensions can be defined \cite{mirlin2000,mirlinRMP08}. The first set uses average moments $\langle P_q \rangle$ giving dimensions $D_q$, and the second uses typical moments $\exp \langle \ln P_q \rangle$, giving dimensions $D_q^{\text{typ}}$ . We have checked that our results
are the same for both sets of dimensions, and we present results only for $D_q$,
mainly for $q=2$; we have checked that the conclusions we draw apply equally to
other values of $q$. 

Our first model consists in a system periodically kicked by a discontinuous linear potential \cite{giraud}.  
Its Hamiltonian, defined on a phase space corresponding to the unit torus, with $p$ the momentum and $q$ the space coordinate, is:
\begin{equation}
  \label{defH}
  H(p,q,t)=\frac{p^2}{2}-\gamma\{ q\} \sum_n \delta(t-n)\ ,
\end{equation}
where $\{q\}$ means the fractional part of $q$, $\gamma$ is a real parameter, and the sum runs over all integers. 

The classical dynamics over one period is given by the map $p_{n+1}=p_n+\gamma \mod 1\ , \; q_{n+1}=q_n+2 p_{n+1} \mod 1\ ,$ where $n$ denotes the number of periods.  For irrational $\gamma$, the dynamics is ergodic. For rational $\gamma$, it can be described as pseudointegrable. In such systems, the iterates of one point accumulate inside surfaces which are of arbitrarily high genus, 
different from the integrable case where the dynamics takes place on tori of genus one.  

The corresponding quantum discrete dynamics 
transforms the wave function at time $n$ noted $\psi^n$ to the one at time $n+1$ through the formula $\psi^{n+1}= U \psi^n$. In an $N$-dimensional Hilbert space, $U$ corresponds to an $N\times N $ matrix with coefficients \cite{giraud}
\begin{equation}
  \label{defU}
  U_{kl}
  =\frac{e^{-2\pi\ic k^2/N}}{N}\frac{1-e^{2\ic\pi\gamma N}}{1-e^{2\ic\pi(k-l+\gamma N)/N}}\ ,
\end{equation}
where $k,l$ are quantum numbers associated to momentum, with 
an effective $\hbar$ equal to $1/(2\pi N)$.
In the results shown below a random version of the model is considered \cite{bogomolny,MGG,MGGG}: $e^{-2\pi\ic k^2/N}$ is replaced by $e^{-\ic \phi_k}$ where $\phi_k$ is a random variable uniformly distributed in $[0;2\pi]$. This allows for more stable results and we have checked that the results are the same as with the usual kinetic term, but with less fluctuations.
For irrational $\gamma$, the eigenvectors of (\ref{defU}) are ergodic in phase space. In contrast, for rational $\gamma=a/b$, 
they  are multifractal in the momentum basis
(\ref{defU}), consisting in $b$ strongly fluctuating structures.
This multifractality is weaker and weaker when $b$ increases. In parallel, spectral statistics follow predictions of Random Matrix Theory for irrational $\gamma$, while for rational $\gamma$ they are intermediate between distributions typical of either chaotic or integrable systems \cite{giraud,bogomolny,map_pseudoint}.
 Thus (\ref{defU}) is often called the intermediate map.

Our second model is the 3D Anderson model \cite{Anderson58}, a tight-binding model of electrons with on-site disorder uniformly distributed in $[-W/2,W/2]$. For this model, it is known that a metal-insulator transition takes place at a disorder value $W_c \approx 16.5$ in the band center. At this critical value, wave functions are known to display multifractality \cite{mirlinRMP08}.

\begin{figure}[ht]
\hspace{0.165cm} \includegraphics[width=0.44\textwidth]{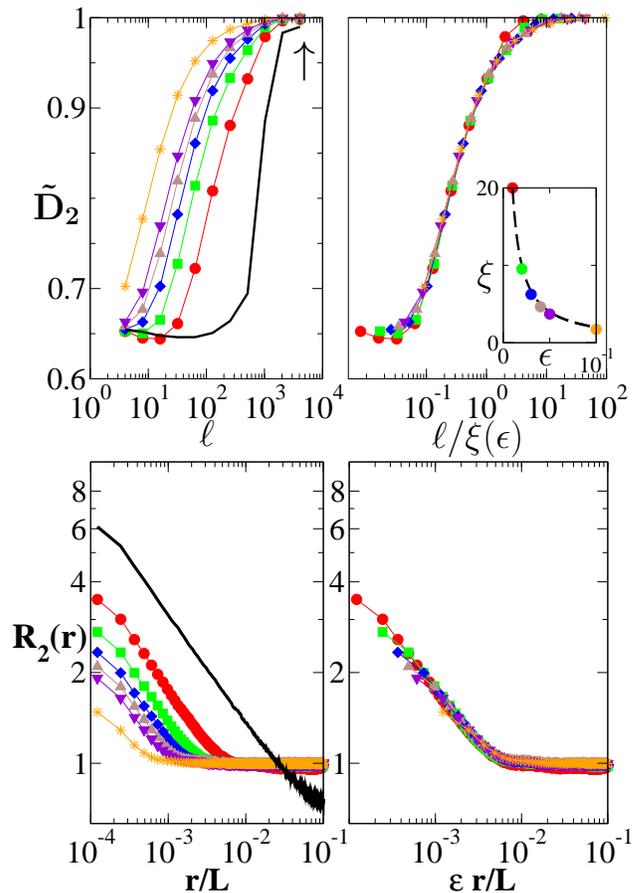}\\
\includegraphics[width=0.46\textwidth]{fig1b.eps}
 \caption{(Color online) Top: Local dimensions $\tilde{D}_2$ (see text) for eigenvectors of (\ref{defU}) with $N=2^{13}$ and $\gamma=1/5$ for smoothing lengths resp. $\epsilon=0$ (black full curve), $0.01$ (red circles), $0.02$ (green squares), $0.03$ (blue diamonds), $0.04$ (brown triangles up), $0.05$ (purple triangles down), $0.1$ (orange stars). Left: raw data for $\tilde{D}_2$ vs the boxsize  (arrow indicates the value $\ell=N$); right: $\tilde{D}_2$ vs the rescaled boxsize $\ell/\xi(\epsilon)$, with $\xi$ normalized as $\xi (\epsilon=1/N)=N/5$, its known value for $\epsilon \rightarrow 0$ (see text). Inset: Numerically obtained scaling length $\xi(\epsilon)$ (circles), black dashed line is the relation $\xi(\epsilon) \propto \epsilon^{-1}$. Bottom: correlation function $R_2(r)$, same parameters and color code as for top. Left: raw data; right: rescaled data using the relation $\xi(\epsilon) \propto \epsilon^{-1}$.}
 \label{D2_FSS_vs_eps}
\end{figure}

We now turn to our results.
The first model (\ref{defH}) has a
discontinuous potential. In many experimental situations, the singularity  
will be smoothed over a certain length.
We model the smoothing by replacing the discontinuous potential  by a $C^1$ function coinciding with $-\gamma\{ q\}$ over $[0,1-\epsilon]$ and equal to a cubic interpolating polynomial over the interval $[1-\epsilon, 1]$. 
We have studied how the multifractality depends on the scale $\ell$ of the coarse-graining in the box-counting method for different values of $\epsilon$. Indeed, in physics one must always be concerned with the ranges in which the scaling $P_{q} \sim (\ell/L)^{D_q (q-1)}$ holds.  For a fixed value of
the moment order $q$ we define a local multifractal dimension
$\tilde{D}_q(\ell,\epsilon)=\frac{1}{q-1} \frac{d\ln  P_q  }{d\ln \lambda}$, where $\lambda=\ell/L$ with $L=N$ the linear size of the system. 
We find that there exists
a characteristic length $\xi(\epsilon)$ below which the local dimensions $\tilde{D}_q$ do
not vary with $\ell$, indicating a true multifractal behaviour.
On the other hand, the local dimensions converge to the ergodic value $\tilde{D}_q =1$ for $\ell \gg \xi(\epsilon)$ (see Fig.~\ref{D2_FSS_vs_eps}, top left). Moreover, we have observed  that all the 
curves for different smoothings $\epsilon$ collapse onto a single one when $\ell$ is scaled with the suitable length $\xi(\epsilon)$, see an example in
Fig.~\ref{D2_FSS_vs_eps} (top) for $q=2$. This shows that the data follow the scaling behavior:
\begin{equation}
  \label{xi_vs_eps}
  \tilde{D}_q(\ell,\epsilon)=G_q\left(\frac{\ell}{\xi(\epsilon)}\right),
\end{equation}
with $G_q$ a scaling function independent of $\epsilon$, and with the scaling parameter $\xi(\epsilon) \propto \epsilon^{-1}$. We have checked that this scaling is valid in the range $1 \gg \epsilon \gg 1/N$.

Another way to illustrate our results consists in considering the $2-$point correlation function $
R_2(r)= N^2 \langle |\psi_i|^2 |\psi_{i+r}|^2 \rangle $,
where the average is performed over both $i$ and the random phases $\phi_k$.
This correlation function is expected to be related to the multifractal dimension $D_2$, via  $R_2(r)\sim  r^{\eta}$ with $\eta=D_2 -1$ for $\frac{r}{L}\to 0\ $, see e.g. \cite{mirlinRMP08}. 
 It is clear that the power-law behavior survives for
$\epsilon>0$ (see Fig.~\ref{D2_FSS_vs_eps} bottom). The main effect of the smoothing is again the
emergence of a characteristic length $\xi(\epsilon)$, above which $R_2(r)$ is not algebraic anymore. When it is algebraic,  we find for both $\epsilon =0$ and $\epsilon >0$ an exponent $\eta \approx -0.36$ in very good agreement with the value $D_2$ extracted from Fig.~\ref{D2_FSS_vs_eps} top for $\ell \rightarrow 0$.  Thus the multifractal fluctuations are left {\em unchanged} below the characteristic length $\xi(\epsilon)$.
We have checked that other ways of smoothing the potential lead to the same conclusions \cite{long}.

A physical interpretation of $\xi(\epsilon)$ for our model is related to the initial shape of the potential in (\ref{defH}). For $\epsilon=0$ the discontinuity of the potential is resolved at a scale
$1/N$,  and the multifractality appears below a scale of order $N/b$ ($b$ coming from the fact that classical structures have $b$ components), which explains why $\tilde{D}_2 \rightarrow 1$ when $\ell \gg N/b$. The smoothing introduces a new effective width $\epsilon$ for the singularity. Hence multifractality in the momentum basis survives below a scale of order $\xi(\epsilon)\propto 1/\epsilon$ for $\epsilon \gg 1/N$ .

This scenario of a characteristic length bounding the scale of the multifractal structure is similar to the one found in Anderson-like transitions when the system is close to the transition point \cite{pietronero,kravcuev,romer}. In this case the relevant characteristic length coincides with the localization length in the insulating phase and with the correlation length in the metallic phase.  The case of the intermediate map can be seen as a multifractal metal described in \cite{kravcuev}. We emphasize that  in this scenario,  multifractality always survives the perturbation at a sufficiently small scale.

\begin{figure}[ht]
  \begin{center}
    \includegraphics[width=0.45\textwidth]{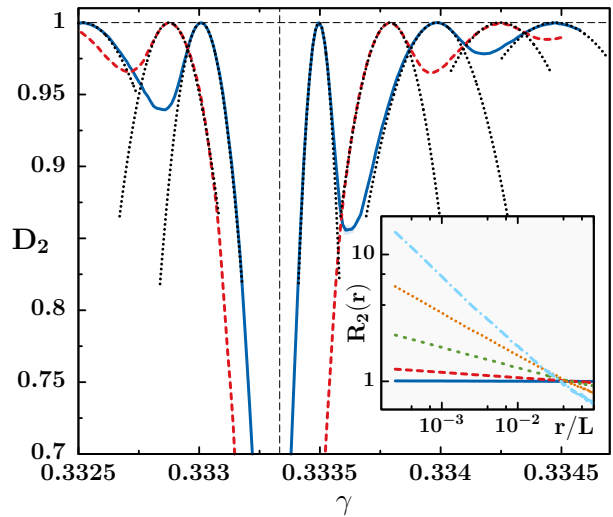}
  \end{center}  
  \caption{(Color online) Dimension $D_2(\gamma)$ for the model (\ref{defU}) in the vicinity of $\gamma=1/3$ for $N=3^7$ (red dashed line), $N=2^{11}$ (blue full line). Black dotted parabolas correspond to Eq.~(\ref{dqfinalrandom}). 
Inset: correlation function $R_2(r)$ for $N=2^{12}$. The curves correspond to $\gamma=1/3+\epsilon/(3N)$ for resp. $\epsilon=0$ (light blue dotted-dashed line), $0.25$ (orange dotted line), $0.5$ (green dashed line), $0.75$ (red long-dashed line), $0.95$ (blue solid line).
}
  \label{Dq_vsg}
\end{figure}

We now turn to the second scenario. A natural perturbation of (\ref{defH}), when $\gamma$ is a
rational value, is to slightly change this value, at fixed $N$. A striking observation that we made is the absence of any characteristic length in the fluctuations of the wave functions. Indeed,  
Fig.~\ref{Dq_vsg} shows the variation of $D_2$ and $R_2(r)$ close to a rational point.  For different $\gamma$ values close to $\gamma=1/3$, the correlation function $R_2(r)$ behaves as a power law in the same
range of $r$: hence there is no characteristic length here. In contrast with the first scenario, the perturbation now induces a change of the algebraic decay of $R_2$, hence a change of $D_2$. The same conclusions can be drawn from $P_2$  (data not shown).

For $N \rightarrow \infty$ one has $D_q=1$ for all irrational values of $\gamma$, but for finite  $N$ the curve will be smoothed out over a certain scale, as
 shown for $D_2$ in Fig.~\ref{Dq_vsg}.  We found that the vicinity of rational values is related to a mathematical model called the Ruijsenaars-Schneider model \cite{ruijsc}.
Using a perturbative approach similar to the one used in \cite{BogGir}, we can predict analytically the behavior of $D_q$ near its local extrema. 
Technical details will be published elsewhere \cite{long} but
the results can be summarized as follows. Around $\gamma=1/b$, local extrema of $D_q$  are
located at $\gamma_k=1/b+(k-s/b)/N$, where $s=N$
mod $b$, and $k=0, \pm 1, \pm 2, \dots$. Around those extrema
the intermediate map shows weak multifractality and for $ |\gamma-\gamma_k|\simeq 1/N$ the multifractal dimension is :
\begin{equation}
  \label{dqfinalrandom}
  D_q\simeq 1- q b \left[\frac{N(\gamma-\gamma_k)}{(k b-s)}\right]^2.
\end{equation}
Note that this theory, in very good agreement with our numerics (see Fig.~\ref{Dq_vsg}), again does not contain any characteristic length.

A similar phenomenon can be seen when the basis is slightly deformed. Indeed  multifractal properties depend on the basis choice, and in experimental implementations the measurement basis cannot be chosen at will.

\begin{figure}[ht]
 \includegraphics[width=0.9\linewidth]{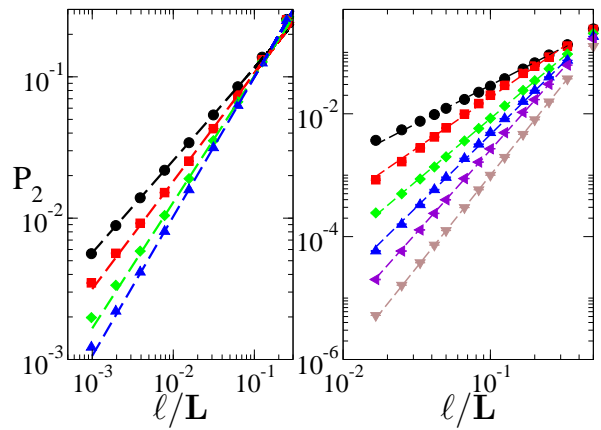}
  \caption{(Color online) Moments $P_2$ vs boxsize for different basis deformation strengths $\epsilon$. Left: map (\ref{defU}) for $N=2^{12}$, $\gamma=1/5$ and for resp. $\epsilon=0.0013$ (black circles), $0.013$  (red squares), $0.02$ (green diamonds), $0.1$ (blue triangles); right: 3D Anderson model at the transition point, of size $N=L^{3}$ with $L=120$ for resp. $\epsilon=t/\tau_{\rm Th}=0$ (black circles), $8.9 \ 10^{-11}$ (red squares), $1.1 \ 10^{-9}$ (green diamonds),  $2.8 \ 10^{-7}$ (blue triangles up),  $8.7 \ 10^{-6}$ (purple triangles tilted),  $2.3$ (pink triangles down).}
  \label{Dq_deformId}
\end{figure} 

We thus investigate the behavior of multifractality for the map (\ref{defU}) under a generic change of basis. The unitary matrix defining the basis change is taken to be
 $ \tilde{U}=\exp (\ic \epsilon H)$, 
where $\epsilon$ is the deformation parameter and $H$ an element of the GOE ensemble of Random Matrices. Moments averaged over the GOE ensemble are plotted for several values of $\epsilon$ in Fig.~\ref{Dq_deformId} (left), showing that the slope
changes with $\epsilon$ at all scales, which corresponds to our second scenario.

\begin{figure}[ht]

  \hspace{0.1cm}  \includegraphics[width=0.44\textwidth]{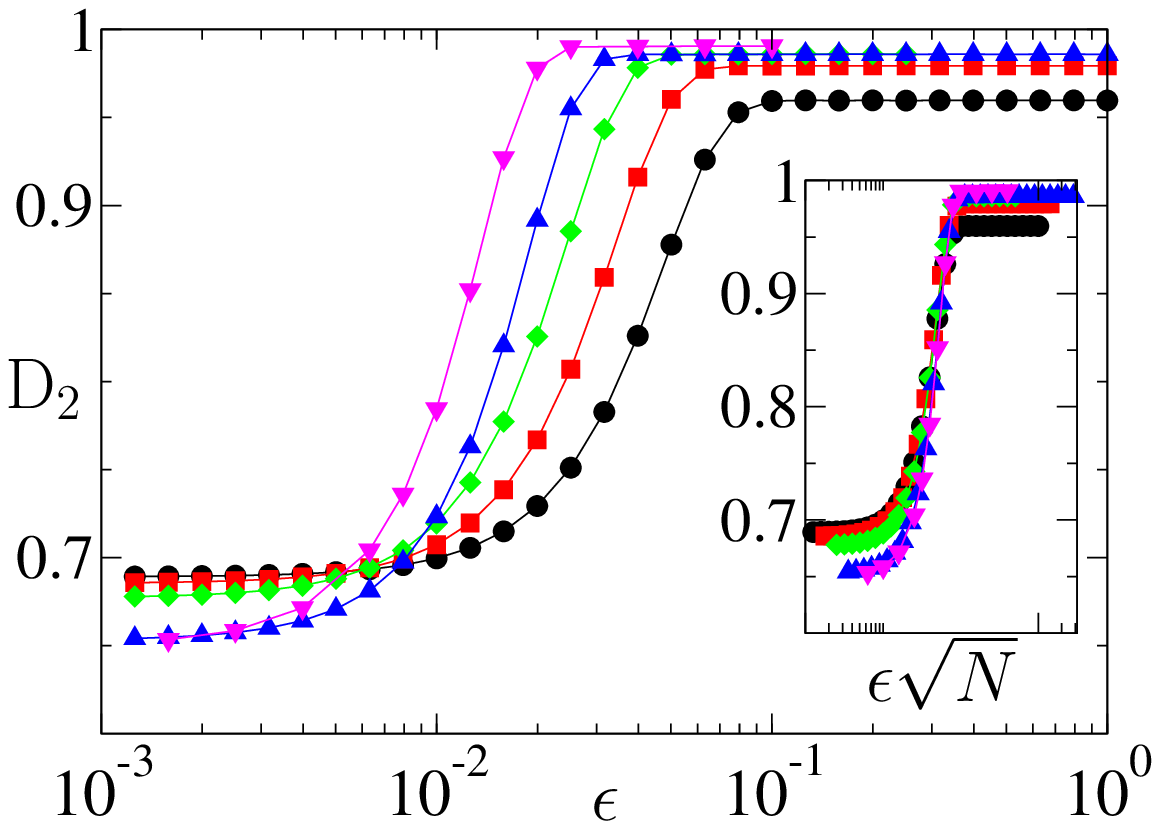}
\includegraphics[width=0.43\textwidth]{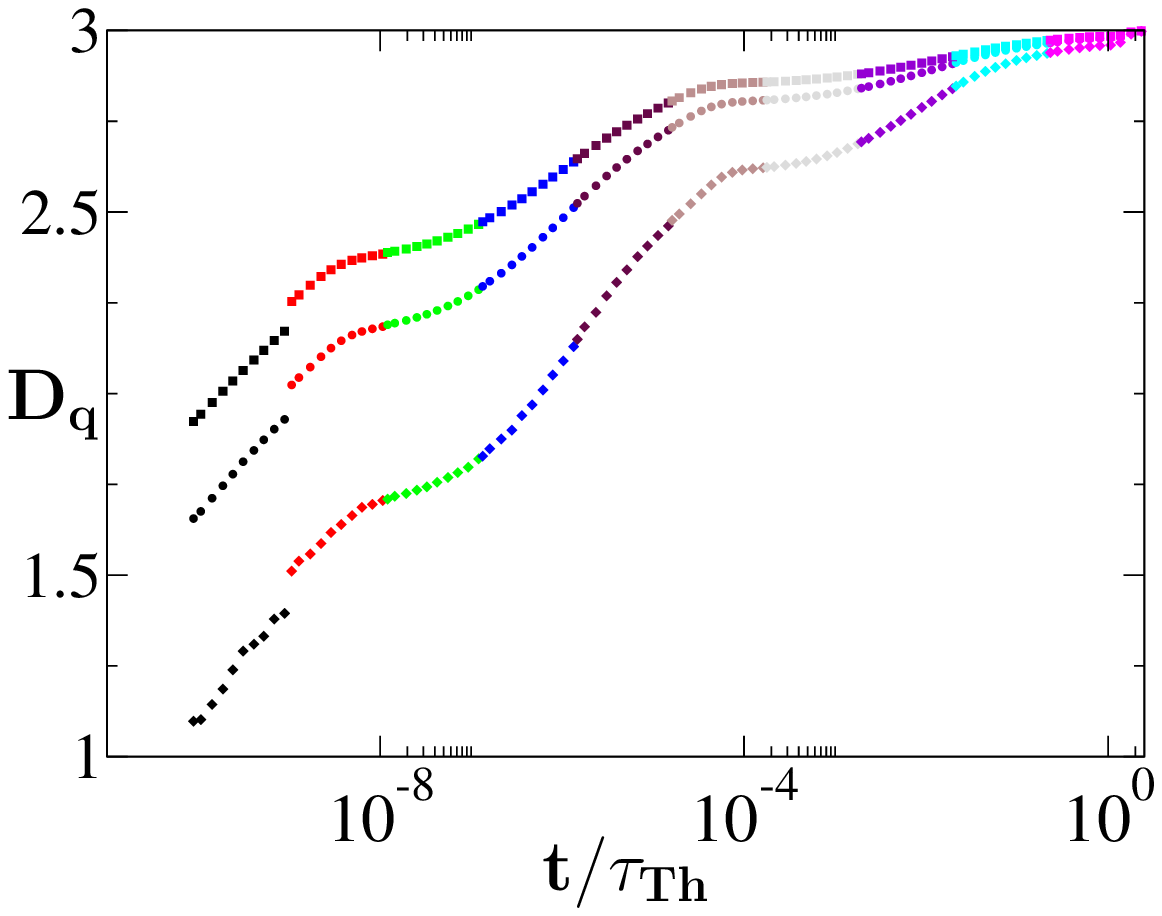}
  \caption{(Color online) Dimensions $D_q$ vs basis deformation strength $\epsilon$. Top: $D_2$ for the map (\ref{defU}), $\gamma=1/5$ for resp.  $N=2^{9}$ (black circles), $2^{10}$ (red squares), $2^{11}$ (green diamonds), $2^{12}$ (blue up triangles), $2^{13}$ (magenta down triangles); inset: theoretical scaling with $\epsilon \sqrt{N}$ (see text); bottom: from top to bottom $D_{1.5}$, $D_2$ and $D_4$ for the 3D Anderson model,  size $N=120^{3}$ (from left to right: resp. $\tau_{\rm Th}=1.19 \ 10^{11}$ (black), $9.3 \ 10^{9}$ (red), $8.3 \ 10^{8}$ (green), $7.5 \  10^{7}$(blue), $6.8 \  10^{6}$ (dark brown),  $6.2 \  10^{6}$ (brown), $5.6 \  10^{4}$ (grey), $5121.2$ (dark purple), $468.5$ (cyan), $43.1$ (magenta).
}
  \label{Anderson}
\end{figure}

This is confirmed by a perturbation theory that we have developed (see \cite{long} for more details).
Upon basis change, a state $\ket{\psi}$ is changed into some state $\ket{\tilde{\psi}}=\tilde{U}\ket{\psi}$. At second order, it reads
$\ket{\tilde{\psi}}=\ket{\psi}+i\epsilon H\ket{\psi}-\frac{\epsilon^2}{2}H^2\ket{\psi}$.
Upon averaging over the GOE ensemble, terms linear in $H$ will vanish in the moments $P_2$, while by independence of GOE matrix entries only quadratic terms of the form $H_{mn}^2$ will survive. 
Thus the moments of $\ket{\tilde{\psi}}$ read
$\sum_n|\tilde{\psi}_n|^4=\sum_n|\psi_n|^4+2\epsilon^2v^2\left(1-\frac12|\sum_n\psi_n^2|^2-\frac{N}{2}\sum_n|\psi_n|^4\right)$, where $v^2$ denotes the variance of the GOE matrix elements (here $v=1$).
Moments are multiplied by an effective factor $1-\epsilon^2 v^2 N$ (assuming that the term in $|\sum_n\psi_n^2|^2$ is negligible), so that multifractality is destroyed for $\epsilon$ of order $1/\sqrt{N}$. This theory, confirmed by our numerics (see Fig. \ref{Anderson}, top), does not single out any scale where the behavior will change, confirming that indeed the moments are modified at all scales by the perturbation.

Remarkably enough, the same behavior appears at the Anderson transition
when the basis is deformed.  Using large-scale numerical simulations \cite{jada}, we have computed the moments of the wave functions (eigenvectors) of the 3D Anderson model for sizes up to $N=L^{3}$ with $L=120$. 
As it was impossible to implement the change of basis as above,  given the size of our matrices, we used instead the evolution operator corresponding to the quasiperiodic kicked rotor $\tilde{H}=p^2/2 + K \cos \theta (1+ \eta \cos (\omega_2 t) \cos (\omega_3t)) \sum_n \delta(t-n) $, with $\eta=0.8$, $\omega_2= 2\pi \sqrt{5}$, $\omega_3 =2\pi \sqrt{13}$ and $\hbar=2.89$. 
This 1D system is known to display an Anderson transition for $K=K_c \approx 4.7$ \cite{cold,dima}. Here we used large values of $K \gg K_c$ to ensure a diffusive dynamics where statistics are known to be close to Random Matrix results.   

In Fig.~\ref{Dq_deformId} (right) we show the moments $P_2$ for various values of the perturbation $\epsilon$ of the basis. The curves are similar at all scales, but with a slope which varies with the perturbation strength. This indicates that the multifractality is affected in the same way at all scales. This leads to a disappearance of multifractality as shown in
Fig.~\ref{Anderson} (bottom).
The perturbation here corresponding to the evolution operator of a diffusive system, a natural scale is the Thouless time $\tau_{\rm Th}$, defined as the ratio $L^{2}/D$, where $D$ is the diffusion constant. $\tau_{\rm Th}$ is the characteristic time where ergodicity sets in. Fig.~\ref{Anderson} (bottom) shows that indeed
this is the relevant scale, the variation of the parameter $K$ of the model enabling us to probe several orders of magnitude of $\tau_{\rm Th}$.
In addition, we checked that the results presented for both models are representative of a real experimental situation where a single basis is imposed by the setup, i.~e.~without averaging over basis change (data not shown). 

In this Letter, we have investigated how the quantum multifractality is modified by various generic perturbations. We have identified two different scenarios. In the first scenario, a characteristic length appears, which bounds the scale of the multifractal fluctuations of the wave functions. In the second scenario, multifractality is destroyed equally at all scales.  Both scenarios are found in the two models we have investigated, which represent the two main classes of systems displaying quantum multifractality: pseudointegrable systems and Anderson-type models at criticality.  From an experimental point of view, in the first scenario one can compensate a finite perturbation by using high resolution to resolve very small scales. On the contrary, in the second scenario one definitely needs to control the perturbation below a critical value.  These results give a theoretical understanding which should provide guidance towards the observation of quantum multifractality in a real experimental setting.

\begin{acknowledgments} We thank Olivier Herscovici for discussions and insights, and one referee for his stimulating remarks. We thank CalMiP for access to its supercomputers,
the FNRS and the University Paul Sabatier for funding (OMASYC project). This work was supported by Programme Investissements d'Avenir under the program ANR-11-IDEX-0002-02, reference ANR-10-LABX-0037-NEXT. IGM received support from ANCyPT grant PICT 2010-1556 
and from CONICET grant PIP 114-20110100048. JM is grateful to the University of Li\`{e}ge for the use of the NIC3 supercomputer (SEGI facility), and for funding (project C-13/86).
JM and IGM received financial support from a CONICET-FNRS bilateral project, and BG and IGM from the CONICET-CNRS bilateral project PICS06303.
\end{acknowledgments}


\begin{thebibliography}{99}
\bibitem{turbulence} C.~Meneveau and K.~R.~Sreenivasan,  Phys. Rev. Lett. {\bf 59}, 1424
(1987); J.-F.~Muzy, E.~Bacry and A.~Arneodo, Phys. Rev. Lett. {\bf 67}, 3515 (1991).
\bibitem{stock} B.~B.~Mandelbrot, A.~J.~Fisher and L.~E.~Calvet,  Cowles Foundation Discussion Paper No. 1164 (1997).
\bibitem{cloud} S.~Lovejoy and D.~Schertzer,
Journal of Geophysical Research {\bf 95}, 2021 (1990).
\bibitem{kohmoto} H.~Hiramoto, M.~Kohomoto, Int. J. of Mod. Phys. B, {\bf 6}, 281 (1992).
\bibitem{mirlin2000}
A.~D.~Mirlin, Phys. Rep. {\bf 326}, 259 (2000).
\bibitem{mirlinRMP08} F.~Evers and A. D. Mirlin, Rev. Mod. Phys. 
{\bf 80}, 1355 (2008).
\bibitem{romer} A.~Rodriguez, L.~J.~Vasquez and R.~A.~R\"omer, Phys. Rev. Lett. {\bf 102}, 106406 (2009);  Eur. Phys. J. B {\bf 67}, 77 (2009); A. Rodriguez, L. J. Vasquez, K. Slevin, and R. A. R\"omer, Phys. Rev. Lett. {\bf 105}, 046403 (2010); A. Rodriguez, L. J. Vasquez, K. Slevin, and R. A. R\"omer, Phys. Rev. B {\bf 84}, 134209 (2011).
\bibitem{huckenstein} B.~Huckestein, Rev. Mod. Phys. {\bf 67}, 357 (1995).
\bibitem{PRBM} A.~D.~Mirlin, Y.~V.~Fyodorov, F.-M.~Dittes, J.~Quezada, and
  T.~H.~Seligman, Phys. Rev. E {\bf 54}, 3221 (1996); J.~A.~Mendez-Bermudez, A.~Alcazar-Lopez and I.~Varga, Europhys. Letters {\bf 98} 37006 (2012).
\bibitem{ossipov} Y.~V.~Fyodorov, A.~Ossipov and A.~Rodriguez, J. Stat. Mech. L12001 (2009).
\bibitem{indians1} N.~Meenakshisundaram and A.~Lakshminarayan,  Phys. Rev. E {\bf 71}, 065303(R) (2005).
\bibitem{indians} J.~N.~Bandyopadhyay, J.~Wang and J.~Gong, Phys. Rev. E {\bf 81}, 066212 (2010).
\bibitem{garciagarcia} A.~M.~Garc\'{\i}a-Garc\'{\i}a and J.~Wang,
Phys. Rev. Lett. {\bf 94}, 244102 (2005).
\bibitem{interm} E.~B.~Bogomolny, U.~Gerland and C.~Schmit,  
Phys. Rev. E {\bf 59}, R1315 (1999); E.~B.~Bogomolny, O.~Giraud and C.~Schmit, Phys. Rev. E
{\bf 65}, 056214 (2002).
\bibitem{wiersig} J.~Wiersig, Phys. Rev. E {\bf 62}, R21 (2000).
\bibitem{giraud} O.~Giraud, J.~Marklof and S.~O'Keefe, J. Phys. A
{\bf 37}, L303 (2004).
\bibitem{bogomolny} E.~B.~Bogomolny and C.~Schmit,
Phys. Rev. Lett. {\bf 93}, 254102 (2004).
\bibitem{MGG} J.~Martin, O.~Giraud and B.~Georgeot,  Phys. Rev. E {\bf 77}, R035201 (2008).
\bibitem{map_pseudoint} E.~B.~Bogomolny, R.~Dubertrand and C.~Schmit,
  Nonlinearity \textbf{22}, 2101 (2009).
\bibitem{MGGG} J.~Martin, I.~Garc\'{\i}a-Mata, O.~Giraud and B.~Georgeot,  Phys. Rev. E {\bf 82}, 046206 (2010).
\bibitem{BogGir} E.~B.~Bogomolny and O.~Giraud, Phys. Rev. Lett. {\bf 106}, 044101 (2011); Phys. Rev. E {\bf 84}, 036212 (2011); Phys. Rev. E {\bf 85}, 046208 (2012). 
\bibitem{MGGG12}  I.~Garc\'{\i}a-Mata, J.~Martin, O.~Giraud and B.~Georgeot,  Phys. Rev. E {\bf 86}, 056215 (2012).
\bibitem{guarneri} I.~Guarneri and G.~Mantica, Phys. Rev. Lett. {\bf 73}, 3379
(1994).
\bibitem{geisel} R.~Ketzmerick, K.~Kruse, S.~Kraut, and T.~Geisel, Phys. Rev.
Lett. {\bf 79}, 1959 (1997).
\bibitem{mantica} G. Mantica, Electronic Transactions in numerical analysis {\bf 25}, 409  (2006).
\bibitem{richard}   A.~Richardella, P.~Roushan, S.~Mack, B.~Zhou, D.~A.~Huse, D.~D.~Awschalom and A.~Yazdani, Science {\bf 327}, 665 (2010).
\bibitem{cold} G.~Lemari\'e, H.~Lignier, D.~Delande, P.~Szriftgiser and J.-C.~Garreau, Phys. Rev. Lett. {\bf 105}, 090601 (2010); M.~Lopez, J.-F.~Cl\'ement, G.~Lemari\'e, D.~Delande, P.~Szriftgiser and J.-C.~Garreau, New J. of Phys. {\bf 15}, 065013 (2013).
\bibitem{cold2} Y.~Sagi, M.~Brook, I.~Almog and N.~Davidson, Phys. Rev. Lett. {\bf 108}, 093002 (2012).
\bibitem{billes} S.~Faez, A.~Strybulevych, J.~H.~Page, A.~Lagendijk and B.~A.~van Tiggelen,  Phys. Rev. Lett. {\bf 103} 155703 (2009).
\bibitem{Anderson58} P.~W.~Anderson Phys. Rev. {\bf 109}, 1492 (1958).
\bibitem{long} R.~Dubertrand, I.~Garc\'{\i}a-Mata, B.~Georgeot, O.~Giraud, G.~Lemari\'e and J.~Martin, to be published.
\bibitem{pietronero} A.~P.~Siesbesma and L.~Pietronero, Europhys. Lett. {\bf 4}, 597 (1987).
\bibitem{kravcuev} E.~Cuevas, V.~E.~Kravtsov, Phys. Rev. B {\bf 76}, 235119 (2007).
\bibitem{ruijsc} S.~N.~M.~Ruijsenaars and H.~Schneider, Ann. Phys. {\bf 170}, 370 (1986).
\bibitem{jada} We used the highly optimized Jadamilu library to diagonalize large sparse matrices, see M. Bollh\"ofer and Y. Notay, Computer Phys. Comm., {\bf 177}, 951 (2007). 
\bibitem{dima} G.~Casati, I.~Guarneri, and D.~L.~Shepelyansky, Phys. Rev. Lett.
{\bf 62}, 345 (1989).
\end{thebibliography}
\end{document}